\documentclass[aps,prl,groupedaddress,twocolumn,showkeys]{revtex4}
\usepackage{graphicx}
\usepackage{amsmath}
\usepackage{verbatim}
\newcommand{\req}[1]{(\ref{#1})}
\newcommand{\be}{\begin{equation}}
\newcommand{\ee}{\end{equation}}
\newcommand{\bea}{\begin{eqnarray}}
\newcommand{\eea}{\end{eqnarray}}

\newcommand{\pr}[1]{\left(#1\right)}
\newcommand{\cro}[1]{\left[#1\right]}

\newcommand{\beas}{\begin{eqnarray*}}
\newcommand{\eeas}{\end{eqnarray*}}

\newcommand{\erf}{{\mathrm erf}}

\newcommand{\bs}{\begin{split}}
\newcommand{\espl}{\end{split}}
\newcommand{\bse}{\begin{subequations}}
\newcommand{\ese}{\end{subequations}}
\newcommand{\ba}{\begin{align}}
\newcommand{\ea}{\end{align}}

\def\erf{\hbox{erf}\,}

\newcommand{\avg}[1]{\langle{#1}\rangle}

\newcommand{\ovl}[1]{\overline{#1}}
\newcommand{\BE}{\begin{eqnarray}}
\newcommand{\EE}{\end{eqnarray}}
\newcommand{\BEn}{\begin{eqnarray*}}
\newcommand{\EEn}{\end{eqnarray*}}
\newcommand{\barr}{\begin{array}}
\newcommand{\earr}{\end{array}}
\newcommand{\bit}{\begin{itemize}}
\newcommand{\eit}{\end{itemize}}
\newcommand{\bc}{\begin{center}}
\newcommand{\ec}{\end{center}}
\newcommand{\ben}{\begin{enumerate}}
\newcommand{\een}{\end{enumerate}}

\begin{document}
\title{Shedding light on El Farol}
\author{Damien Challet}
\affiliation{Theoretical Physics, Oxford University, 1 Keble Road,
  Oxford OX1 3NP, United Kingdom}
\email{challet@thphys.ox.ac.uk}
\author{M. Marsili}
\affiliation{International Center for Theoretical Physics, Strada Costiera 11, 34014 Trieste, Italy}
\email{marsili@ictp.trieste.it}
\author{Gabriele Ottino}
\affiliation{D\'epartement de Physique, Universit\'e de Fribourg,
  P\'erolles, 1700 Fribourg, Switzerland}
\date{\today}

\begin{abstract}
  We mathematize El Farol bar problem and transform it into a workable
  model. In general, the average convergence to optimality at the
  collective level is trivial and does not even require any
  intelligence on the side of agents. Secondly, specializing to a
  particular ensemble of continuous strategies yields a
  model similar to the Minority Game. Statistical physics of
  disordered systems allows us to derive a complete understanding of
  the complex behavior of this model, on the basis of its phase
  diagram.
\end{abstract}
\maketitle

\section{Introduction}

Statistical mechanics has developed powerful tools to tackle
analytically disordered systems with many degrees of freedom. These
tools were recently shown to be applicable to systems of inductive
heterogeneous agents such as the Minority
Game~\cite{CZ97,web,CMZe00,CoolenBatch}. Even if the latter is inspired by
El Farol's bar problem~\cite{Arthur}, the literature on these two
models are rather separate. In particular what the MG has brought to
the understanding of the El Farol problem as defined by Arthur is not
clear. Here we show that all results known about the MG directly apply
to the El Farol problem. 

In the El Farol's bar problem, $N=100$ customers have to decide
independently whether to go or not to the bar, which has a capacity of
$L=60$ seats, the resource level. Attending when the bar is crowded is
not enjoyable.  Customers are inductive rational agents. They use
simple predictor rules, based on the past attendances, to predict
whether the bar will be crowded or not, and behave accordingly.

One important issue is whether the customers, that do not communicate
with each other, are able to synchronize their actions so that the
attendance $A$ is on average equal to the resource level $L$. The main
result of Arthur is that agents need not be endowed by a sophisticated
deductive rationality in order to synchronize. Even inductively
rational agents can do. This results is probably responsible for the
large interest his model aroused~\cite{Casti}. Here we show that even
inductive rationality is not necessary because even zero-intelligence
agents, acting as simple automaton, are able to self-organize to the
comfort level.  The convergence of the average attendance to the
comfort level $L$ is trivial under very generic and reasonable
conditions. The really non-trivial question is whether agents are able
or not to reduce stochastic fluctuations of the attendance $A$ around
$L$. The Minority Game was introduced~\cite{CZ97} exactly to address
this question, though in a simplified model. In what follows, by
focusing on a particular ensemble of strategies for the El Farol bar
customers, we derive a model for which we can use all the machinery
used in the theory of Minority Games to derive a complete picture of
the El Farol bar problem.

The main results are that there is an optimal complexity of the
strategies which agents should consider, depending on their number
$N$.  More precisely, for a fixed ratio $\ell=L/N$ of seats to agents,
coordination is optimal when agents use predictor strategies based on
$m\approx \alpha_c \log_2 N$ values of the past attendance. If $m\ll
\alpha_c \log_2 N$ crowd effects occur whereas when $m\gg \alpha_c
\log_2 N$ the information on past attendance is way too redundant. 

We also show the importance for the agents to use consistent
predictors. On the one hand, a small systematic bias affects
considerably the results. On the other, an El Farol problem with
consistent strategies is equivalent to a Minority Game. This implies that  a
large part of literature on the latter model is also directly relevant
to the study of the former without any modification. Inconsistent
strategies correspond to biased strategies, or equivalently to a tunable
resource level, in Minority Games. Such issues were investigated numerically in
Refs.~\cite{JohnsonBias,JohnsonAsym,cara} with numerical simulations.

\section{A mathematical formalism for El Farol bar problem}

In Arthur's paper, each customer uses the public knowledge of 
past $m$ weeks' attendance 
\[
{\cal I}_t=\{A(t-m),A(t-m-1),\cdots,A(t-1)\}
\]
in order to determine whether to go the the bar, or to stay at home.
She crafts a prediction of the next attendance $A(t)$. If her
prediction is larger than the resource level, she stays at home, else
she goes to the bar. The learning procedure is inductive: she has a
personal set of $S$ fixed strategies based on $S$ different attendance
{\em predictors} ${\cal A}_{i,1},\cdots,{\cal A}_{i,S}$. These are
functions mapping the information ${\cal I}$ about the past $m$
attendances to the integer prediction ${\cal A}_{i,s}({\cal
I})\in\,[0,N]$ of the next attendance $A(t)$.

Each attendance predictor ${\cal A}_{i,s}$ $s=1,\cdots,S$ 
 should not be rewarded depending on their precision 
but rather on the payoff they give to an agent who follows their advice.
So if $A(t)=59$, a prediction of 5 is better that a prediction of 61. More precisely, every predictor $s$ has a
score ${\cal U}_{i,s}$ associated to it, that evolves according to 

\[
{\cal U}_{i,s}(t+1)={\cal U}_{i,s}(t)
+\Theta\{[{\cal A}_{i,s}({\cal I}_t)-L][A(t)-L]\}, 
\]

\noindent
where $\Theta(x)$ is the Heaviside function [$\Theta(x)=0$ for $x<0$
  and $\Theta(x)=1$ for $x\ge 0$],
\[
A(t)=\sum_{i=1}^N a_i(t),
\] 
is the attendance at time $t$ and $a_i(t)$ is the choice of customer 
$i$ at time $t$, which is determined by
following her best predictor at that time.
Mathematically, 
\[
s_i(t)=\arg \max_{s'}{\cal U}_{i,s'}(t)
\]
and 
\[
a_i(t)=\Theta[L-{\cal A}_{i,s_i(t)}({\cal I}_t)].
\]
This is the setup of the game proposed by Arthur~\footnote{In practice one
should consider a non integer $L$, else the system can
freeze artificially at $A=L$.}. He did not specify what the predictor
space was, but gave only a few examples of predictors, such as the
average of the last attendance numbers ${\cal A}({\cal
  I}_t)=1/m\sum_{t'=1}^m A(t-t')$, or a mirror number of the
attendance of $t-3$, i.e., ${\cal A}({\cal I}_t)=L-A(t-3)$. Fogel {\em et
al}~\cite{Fogel} used auto-regressive functions ${\cal A}({\cal
  I}_t)=\sum_{k=1}^m f_{k}A(t-k)$, where $f_k$ are real numbers.

Specifying the ensemble of predictors from which agents draw
strategies is a key issue in the El Farol bar problem.

\subsection{Predictors and strategies}

In general, a predictor is a function from ${\cal I}\in [0,N]^m$ to
${\cal A} \in [0,N]$. There are $(N+1)^{(N+1)^m}$ different such
functions. A {\em strategy} instead, is a function $a({\cal I})$ from
the set of possible informations ${\cal I}\in [0,N]^m$ to an action
$a\in \{0,1\}$. Each strategy can be be considered as the result of
the prescription of a predictor: $a({\cal I}_t)=\Theta[L-{\cal
A}({\cal I}_t)]$. 

However, simple counting shows that there are $2^{(N+1)^M}$ possible
strategies, which is way smaller than the number of predictors for
large $N$. Hence many different predictors ${\cal A}({\cal I})$
correspond to the same strategy $a({\cal I}_t)$. For a particular
strategy $a({\cal I})$ with $\sum_{\cal I}a({\cal I})={\cal K}$, there
are
\be
{\cal N}(a)=(N-L)^{\cal K} L^{(N+1)^M-{\cal K}}
\label{calN}
\ee
predictors ${\cal A}$ which are consistent with that strategy. 

Eq. \req{calN} implies that not all strategies are equivalent, in
principle. In order to illustrate this point, let us consider the case
of a strategy resulting from a predictor taken at random. When $N,L\gg
1$, we almost surely pick a strategy which prescribes to go a fraction
$L/(N+1)$ of times. Indeed by Eq. \req{calN}, almost all strategies
have this property.

As a byproduct we see that Arthur's result that the attendance
self-organizes to the comfort level $L$ is trivial if agents draw
predictors uniformly and at random from the whole predictor space. The
attendance self-organizes to the comfort level $L$ for the simple
reason that agents following predictor strategies will attend with a
probability $L/(N+1)$.

Actually it is desirable to restrict the ensemble of predictors from
which agents draw, to those having some consistency and continuity
properties. Consistency means that the predictor should be consistent
with past observations: for example if $A(t-k)$ fluctuates around some
value $L$ a consistent predictor would also have ${\cal A}\approx L$.
A predictor of the Fogel type ${\cal A}({\cal I})=\sum_{k=1}^M f_k
A(t-k)$ should be such that $\sum_{k=1}^M f_k =1$, else it would
predicts systematically an attendance which is larger than the true
one, hence, not be consistent.

A minimal requirement of consistency is that the resulting strategies
be unbiased, i.e. that ``on average'' they prescribe to attend a
fraction $L/(N+1)$ of the times. A mathematical formalization of this
property entails non-trivial considerations and it will not be pursued
here\footnote{It is not even clear that such a formalization is
possible {\em a priori}. Notice that ``on average'' in the discussion
above refers to a probability distribution on ${\cal I}$ which is that
generated by the game's dynamics itself.}. Rather we shall later
introduce explicitly the bias of strategies as an external parameter
and study the collective behavior as a function of it.

Strategies derived from a random predictor, are unbiased but fail to
have a minimal degree of ``continuity''. Loosely speaking, continuity
of ${\cal A}$ means that if the change in the information ${\cal I}$
is small, the change in the prediction ${\cal A}({\cal I})$, or at
least in the prescribed action $\Theta[L-{\cal A}({\cal I})]$ should
also be small (or rare). For example, the action prescribed by the
strategy should change only rarely when the attendance of a past day
changes by a small amount. Considering that past attendance may also
be subject to observation errors, continuity is a quite desirable
robustness property of strategies.

\subsection{A workable model}

In the following we focus on a particular ensemble of strategies,
which is obtained by reducing the information space. The intuition is
that what is really telling about the value of a past attendance
$A(t-k)$ is whether that was below or above the comfort level $L$,
which is a binary information. In other words we consider strategies

\be 
a_{s,i}^{\mu}=\Theta[L-{\cal A}_{s,i}({\cal I})]
\label{asimu}
\ee
which depend only on the information
\be
\mu(t)=\{\Theta[L-A(t-1)],\ldots,\Theta[L-A(t-m)]\}.
\label{mut}
\ee 
Clearly strategies derived in this way have a high degree of
continuity. 

Not all strategies derived by a predictor $a_{s,i}^{\cal
I}=\Theta[L-{\cal A}_{s,i}({\cal I})]$ are of the form described
above. There are only $2^{2^M}$ strategies of this type, which means
that a reduction of the information space also implies a strong
reduction of the strategy space~\footnote{In particular, a randomly
drawn ${\cal A}$ almost surely leads to a strategy which has not this
form.}.

The fact that agents use a strategy space whose size is independent of
$N$ makes much sense. In context where agents interact with a crowd,
their behavior is insensitive to the exact size $N$ of the population.

Henceforth we assume that each agent is assigned $S$ strategies
randomly drawn from this pool. More precisely, for each $i,s$ and
$\mu$ we draw $a_{s,i}^\mu$ independently from the distribution
\[
P(a)\equiv{\rm
Prob}\{a_{s,i}^\mu=a\}=\bar a\delta(a-1)+(1-\bar a)\delta(a).
\]
As explained above, the induced strategy space of the El Farol problem
is such that on average, agents attend the bar with a frequency
$L/N$. In other words, binary strategies of the El Farol bar problem
account {\em a priori} for the convergence of the attendance to $L$ on
average.  Hence we shall consider below an ensemble of strategies such
that the average of $a_{s,i}^\mu$ is $\bar a\approx L/N$.  Actually we
shall see that small deviations of $\bar a$ from $L/N$ may have a large
effect in the limit $N\to\infty$.

A further simplification, which does not change the qualitative
nature of the results~\cite{MC01} amounts to consider a
linear dynamics of the strategy scores

\be
\label{payoffMG}
U_{i,s}(t+1)=U_{i,s}(t)-(2a_{i,s}^{\mu(t)}-1)[A(t)-L].  
\ee 

Note that
the strategies that predict the correct choice are rewarded whereas
those prescribing a wrong choice are punished.

The understanding of the behavior of this model is made complex by the
feedback of the fluctuations of $A(t)$ with the dynamics of
information $\mu(t)$. Notice that Eq. \req{mut} is equivalent to
assuming that $\mu(t)$ follows the non-linear dynamics
\be
\mu(t+1)=\left|2\mu(t)+\Theta[L-A(t)]\right|_{2^m} 
\label{trueinfo}
\ee

\noindent
where $|\ldots|_P$ is the modulus $P$ operation. The behavior of the
Minority Game is largely unaffected if this dynamics is replaced by a
random draw of $\mu(t)$ \cite{Cavagna,CM99}. As we shall see this is
not the case in our case. Still it is very helpful to introduce a
variation of the model with {\em random information}, where $\mu(t)$
is just randomly drawn, with uniform probability, from the integers
$1,\ldots,P=2^m$ (if the history is random, $P$ can be any integer
number). This is because the model with random information can be
understood in detail within a theory such as that developed for the
MG. This is a quite useful intermediate step toward understanding the
behavior of the El Farol bar problem with true information. In
addition it is also possible to quantify the effects of the dynamics
of true information Eq. \req{trueinfo} along the lines of Ref.
\cite{CM99}.

\begin{figure}
    \centering
    \includegraphics[width=0.4\textwidth]{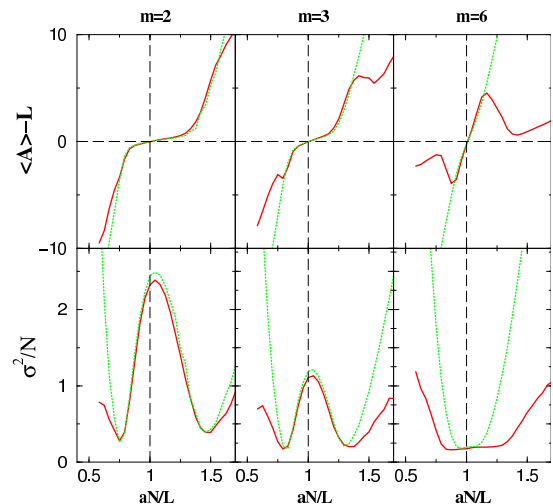}
    \caption{Behavior of the average attendance (top) and of the
    fluctuations (bottom) in the El Farol bar problem with $L=60$
    seats, $\bar a=1/2$ and $m=2,\,3$ and $6$ from left to right.  In
    both cases, the solid (dotted) line refers to true (random)
    information.}
    \label{fig1}
\end{figure}

In order to illustrate the behavior of the model, Fig. \ref{fig1}
shows the results of simulations with $L=60$, $\bar a=1/2$ and
$m=2,3,6$ (from left to right) fixed as a function of $N$. This shows
what happens in a system where the ``environment'' ($L$) and the
adaptive capabilities of agents ($\bar a$ and $m$) are kept fixed,
while the number $N$ of agents increases. The top graph shows the
deviation $\avg{A}-L$ of the average attendance from the comfort
level. As we see when $N\bar a \approx L$ the attendance converges to
the comfort level $\avg{A}\approx L$. However, for small $m$ there is
a whole interval around $N\bar a = L$ where agents are still able to
coordinate efficiently $\avg{A}\approx L$. For $m=2$, the results are
qualitatively the same with true and random information. For larger
values of $m$ the region where $\avg{A}\approx L$ shrinks. In
addition, while $\avg{A}$ maintains a monotonic behavior with random
information, it develops a maximum and minimum for intermediate values
of $\bar a N/L$ beyond which the behavior with true information
markedly departs from that with random information.

It is important to quantify the model's behavior also beyond the
properties of $\avg{A}$. Indeed, we have seen that convergence to the
comfort level is a trivial result in the limit $N\to\infty$. It is a
built-in property of the model which arises from the requirement of
unbiased strategies ($\bar a\approx L/N$). 

The non-trivial cooperative behavior of this system, as forcefully
remarked by the literature on the Minority Game, lies in how and
whether agents manage to decrease fluctuations of the attendance
$A(t)$ around the comfort level $L$. Indeed even if $A(t)$ equals $L$
on average, the distance $|A(t)-L|$ measures the amount of wasted
resources, either unexploited $A(t)<L$ or over-exploited
$A(t)>L$. Therefore, the quality of the cooperation among agents is
measured, at a finer level, by the fluctuations around the resource
level, defined as

\be
\sigma^2=\avg{(A-L)^2}
\ee
where $\avg{\ldots}$ is the average on the stationary state. Given
that $A(t)$ is the sum of $N$ contributions, it is natural to study
the quantity $\sigma^2/N$ which, as we shall see has a finite limit
value in the limit $N\to\infty$.

The behavior of $\sigma^2/N$ is shown in Fig. \ref{fig1}. Away from
the point $\bar a N =L$, the increase of $\sigma^2/N$ is mainly due to
the deviation $\avg{A}-L$ and as before, it differs in the cases of
true and random information. For small $m$, $\sigma^2/N$ displays a
maximum at $\bar a N=L$ which becomes shallower as $m$ increases and
it disappears for $m=6$. This non-trivial behavior suggests that a
small bias in the strategies, of either sign, is beneficial as it
decreases the fluctuations (see also~\cite{JohnsonBias}). 

\begin{figure}
    \centering
    \includegraphics[width=0.4\textwidth]{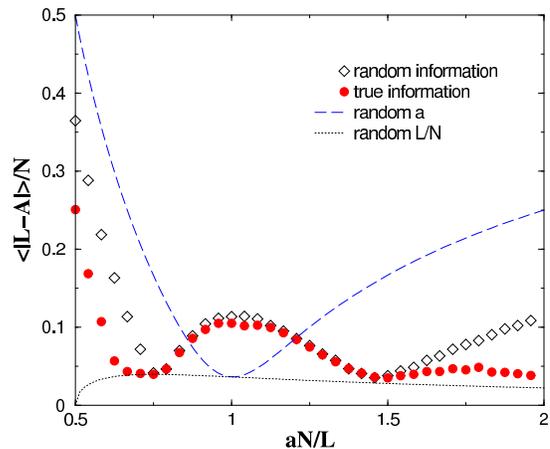}
    \caption{Fraction of losers in the El Farol bar problem with
    $L=60$ seats, $\bar a=1/2$ and $m=2$ for random ($\diamond$) and true
    ($\bullet$) information. The results for a population of random
    agents who attend the bar independently with probability $\bar a$
    and $L/N$ is also shown for comparison (dashed and dotted
    lines).}
    \label{losers}
\end{figure}

This is evident from Fig. \ref{losers}, where we compute the fraction
$\avg{|A-L|}/N$ of unsatisfied agents and compare it with the behavior
of agents who attend the bar at random, either with probability $\bar
a$ or with probability $L/N$. Here we clearly see that in the region
$\bar a N\approx L$ adaptive agents behave less efficiently than
random agents. This effect is related to the emergence of fluctuations
and is stronger for small values of $m$. Efficiency increases for
larger values of $m$.

We shall devote the rest of this paper to explain the non-trivial
behavior displayed by $\avg{A}$ and $\sigma^2$.  The first step will
be extending the analytic approach of Refs.~\cite{CMZe00,MC01} to the
model with random information, which is essentially equivalent to a
Minority Games with biased strategies and tunable resource level.
Then we shall analyze the case with true information.

\section{Statistical mechanics of the El Farol bar problem with 
random information}

Following Ref~\cite{MC01}, one deduces that agent $i$ ends up playing
strategy $s$ with frequency $f_{i,s}$ that minimizes the
quantity
\[
H=\sum_{\nu=1}^P \rho^\nu
(\avg{A|\nu}-L)^2
\]
\noindent
where $\rho^\nu={\rm Prob}\{\mu(t)=\nu\}=1/P$ and 
\[
\avg{A|\nu} =\sum_{i=1}^N \sum_{s_i=1}^S f_{s,i} a_{s,i}^\nu
\]
is the average of $A(t)$ conditional on the event
$\mu(t)=\nu$. This result can be obtained in a straightforward manner
by taking the average of Eq. \req{payoffMG} in the stationary state,
and comparing the resulting equations with the first order conditions
of the minimization of $H$ with respect to $f_{s,i}$.

The function $H$  measures the predictability in the  system, i.e. the
amount of useful information  about the fluctuations of the attendance
which is left in the  signal $\mu(t)$. Indeed if e.g. $\avg{A|\nu}\neq
L$, the signal $\mu(t)$ carries information which is useful to predict
whether one  should attend  or not to  the bar when  $\mu(t)=\nu$. The
fact that the  stationary state corresponds to minimal  $H$ means that
agents exploit to their best the system's predictability.

In terms of statistical mechanics, $H$ can be considered as a
Hamiltonian and its minima can be studied with standard methods. As
long as $H>0$, the stationary state is unique, and the replica trick~\cite{Dotsenko}
gives exact results~\cite{CMZe00,MC01,CoolenBatch}. For the sake of
simplicity, we will focus on the $S=2$ case (see Ref. \cite{MCZe00} for
a generalization to $S>2$). The details of the calculus are of no
special interest, as they mostly replicate previously published
calculations~\cite{CMZe00}.

We shall consider the thermodynamic limit $N\to\infty$ with
\[
\ell\equiv\frac{L}{N}~~~\hbox{and}~~~\alpha=\frac{P}{N}
\]
fixed~\cite{Savit}. Furthermore, in order to study the effect of a deviation of
$\bar a$ around $\ell$ we introduce the convenient parameter $\gamma$
with the equation:

\be\label{gammadef}
\bar a=\ell+\sqrt{\frac{\ell(1-\ell)}{P}}\gamma
\ee

\noindent
and we shall consider $\gamma$ finite in the limit $N\to\infty$. 
As we shall see this is a non-trivial limit. 
For example, the case where $\bar a-\ell\approx
O(1)$ is finite, is trivial because then each agent uses just one
strategy, that which prescribes to go more (less) often if $\bar
a<\ell$ ($\bar a>\ell$) and there is no dynamics at all. 
Eq. \req{gammadef} implies that we consider small
deviation of $\bar a=\ell+O(1/\sqrt{N})$ from the correct value
$\ell=L/N$. Even such a small deviation, which vanishes as
$N\to\infty$, has a finite effect on the global behavior as we shall
see.
At any rate,
other limits can be obtained analyzing the cases where either $\ell$,
$\alpha$ or $\gamma$ vanish or diverge, in the analysis that follows.

After some routine calculations, we find that the predictability $H/N$
is given by

\be\label{H}
H=N\frac{\sqrt{\ell(1-\ell)}}{2}\,
\frac{1+Q(\zeta)+2\gamma^2/\alpha}{[1+\chi(\zeta)]^2}
\ee 

\noindent
and the fluctuations, as long as the stationary state is unique (see 
later), are equal to

\be
\sigma^2=H+N\frac{\sqrt{\ell(1-\ell)}}{2}[1-Q(\zeta)]+N\Sigma.
\label{sigmasq}
\ee

\noindent
In these two equations, $Q$ and $\chi$ are given by

\bea
Q(\zeta)&=&1 - \sqrt{\frac{2}{\pi}} \frac{e^{-\zeta^2/2}}{\zeta}\nonumber- \pr{1-\frac{1}{\zeta^2}}  \erf \pr{\frac{\zeta}{\sqrt{2}}}\\
\chi(\zeta)&=&\cro{\frac{\alpha}{\erf(\zeta/
    \sqrt{2})}-1}^{-1}\label{chi} 
\eea

\noindent
 whereas the parameter $\zeta$ is
uniquely determined by the transcendental equation
\be\label{zeta} 
 \frac{\alpha}{\zeta^2} -Q(\zeta)-1-2\frac{\gamma^2}{\alpha}= 0 
\ee

\noindent
as a function of $\alpha$ and $\gamma$. Finally $\Sigma$ in
Eq. \req{sigmasq} is a term which arises from collective fluctuations
and its calculation requires, in principle, a detailed theory of the
stochastic dynamics of the model (see Ref. \cite{MC01}) which we shall
not pursue here, but whose importance is discussed below.

\begin{figure}
    \centering
    \includegraphics[width=0.4\textwidth]{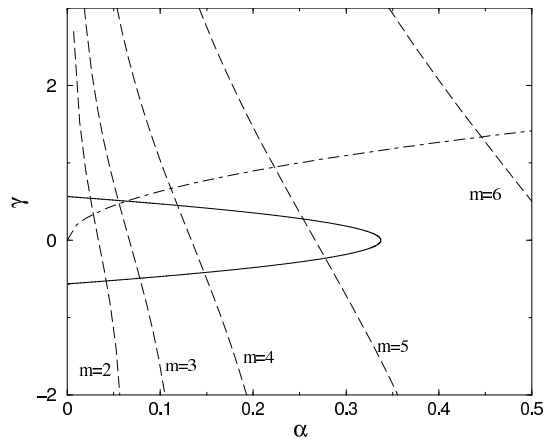}
    \caption{Phase diagram of the El Farol bar problem. The dashed
    lines correspond to the trajectories of systems with $L=60$, $\bar
    a=1/2$ and $m=2,\ldots,6$ as the number of agents increases (from
    bottom to top). The dot-dashed line corresponds to a typical
    trajectory of a system with fixed $L,~N$ and $\bar a>L/N$ as the
    agents' memory changes.}
    \label{alphacgamma}
\end{figure}

When $\gamma=0$, these equations are identical to those which describe
the MG behavior. We briefly recall the resulting picture: When
$\alpha$ is large, the system is in an information rich phase with
positive predictability $H>0$. The predictability $H$ decreases as
$\alpha$ decreases. This can be understood by observing that, at fixed
$P=2^M$, a decrease in $\alpha$ means that the number of agents
increases, and hence their ability to exploit the information. At a
critical value $\alpha_c=0.3374\ldots$ the predictability vanishes and
a phase transition to a symmetric phase with $H=0$ takes place.  The
fact that $H$ measures an asymmetry in the signal means that the phase
transition is related to symmetry breaking~\cite{CM99}. The phase
transition is signaled by the divergence of the spin susceptibility
$\chi$, which is infinite in the whole symmetric phase. The stationary
state is unique, independent of initial conditions, for
$\alpha>\alpha_c$ and these facts conspire~\cite{MC01} in such a way
that $\Sigma\cong 0$ for $\alpha>\alpha_c$. On the contrary, for
$\alpha<\alpha_c$, the stationary state is not unique but it depends
on the initial conditions. Then $\Sigma>0$ can be computed within a
very accurate self-consistent approximation~\cite{MC01}. In particular
this shows that $\Sigma\propto 1/\alpha$ for small $\alpha$.

Therefore, the El Farol bar problem, with the strategy ensemble
studied here, and the MG have the same behavior of fluctuations when
$\gamma=0$, that is, when $N\bar a=L$. When $\gamma\ne0$ the
picture changes in the following way: First we observe that all
quantities depend on $\gamma^2$, then it is enough to consider only
the case $\gamma>0$ since all conclusions extend directly to the case
$\gamma<0$. When $\gamma$ is small, we still have a phase transition
at the point $\alpha_c(\gamma)={\rm erf}(\zeta_c/\sqrt{2})$ where
$\chi\to\infty$. The parameter $\zeta_c$ is the solution of 

\be
\label{zetac} \frac{\erf(\zeta/\sqrt{2})}{\zeta^2}
-Q(\zeta)-1-\frac{2\gamma^2}{\erf(\zeta/\sqrt{2})}= 0 
\ee
\noindent
Figure~\ref{alphacgamma} plots the phase diagram of the game. The
critical line separates the asymmetric phase ($H>0$) from the
symmetric phase ($H=0$). It crosses $\gamma=0$ at
$\alpha_c(0)=0.3374\ldots$, the critical point of the standard
MG~\cite{CMZe00,CoolenBatch}; when $\gamma$ increases $\alpha_c$
decreases and $\alpha_c=0$ for $\gamma\ge 1/\sqrt{\pi}$.

The meaning of the phase diagram is clear: Indeed $H=0$ implies
$\avg{A}=L$.  The symmetric phase is the region of parameters where
the average attendance converges to the comfort level. This region is
also characterized by large collective fluctuations $\sigma^2$ and by
a dependence on initial conditions; in particular, the
fluctuations decrease if the difference of strategy {\em a priori}
valuation increases as discussed in the
MG literature~\cite{MC00,Oxf2,CoolenBatch,MC01}. On the contrary, there is no
equality between $\avg{A}$ and $L$ in the asymmetric phase if
$\gamma>0$.

\begin{figure}
    \centering
    \includegraphics[width=0.4\textwidth]{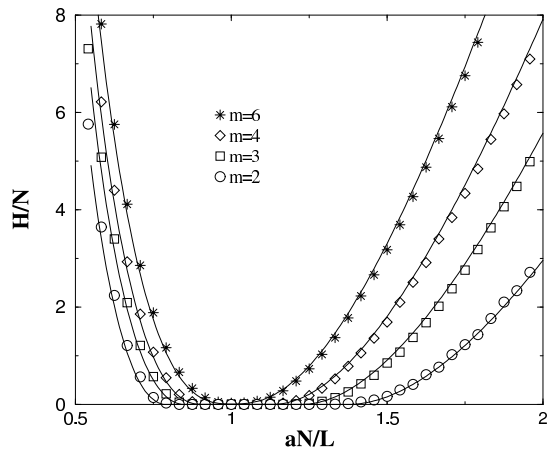}
    \caption{$H/N$ in the El Farol bar problem with true information
    and $L=60$, $\bar a=1/2$ and $m=2,3,4$ and $6$ as a function of
    $N$. Theoretical results (full lines) fully agree with numerical
    simulations (symbols).}
    \label{figH}
\end{figure}

These results explains the complex behavior reported in
Fig. \ref{fig1}. Indeed as $N$ varies with $L,\,\bar a$ and $m$ fixed,
the system follows the trajectories shown in
Fig.~\ref{alphacgamma}. For small values of $m$ these cross the
symmetric phase in the region $\bar a N\approx L$. Fig. \ref{figH}
indeed shows that $H$ computed along these trajectories fully agrees with the
theoretical results (for random information). The symmetric phase is
characterized by large fluctuations, mainly due to dynamic
fluctuations (the term $\Sigma$). This explains the non-monotonic
behavior of $\sigma^2$ in Fig. \ref{fig1} for small values of $m$.
When $m$ increases, the trajectory in Fig.~\ref{alphacgamma} moves
toward larger values of $\alpha$ and, for $m>m_c$, it remains all in the
asymmetric phase. Then $\Sigma=0$, which means that $\sigma^2$ displays
a single minimum at $\bar a N=L$ (i.e. $\gamma=0$).

In spite of the fact that the theory is derived in a particular
$N\to\infty$ limit, our results show that it reproduces accurately
results for moderately small values of $N$. At the same time, it
clearly predicts how the collective behavior depends on the parameters
$N,\,L,\bar a$ and $m$. Any ``experiment'' where one of these
parameters is changed, corresponds to a precise trajectory in the
$(\alpha,\gamma$) phase diagram and a corresponding collective
behavior. For example, for fixed $m$ and $\ell=L/N$, anomalous
fluctuations will arise in an interval of size $1/\sqrt{P}=2^{-m/2}$
as $\bar a$ changes around $\ell$ along a vertical trajectory in
Fig.~\ref{alphacgamma}. The memory
size controls fluctuations. Indeed generally, as $P$ increases,
keeping all other parameters fixed, the systems moves away from the
symmetric phase (see dot-dashed line in Fig.~\ref{alphacgamma}). 

It is precisely at the boundary of the two phases that coordination is
most efficient. This means that there is an intermediate memory length
which is optimal for the collective behavior. 


\begin{figure}
    \centering
    \includegraphics[width=0.4\textwidth]{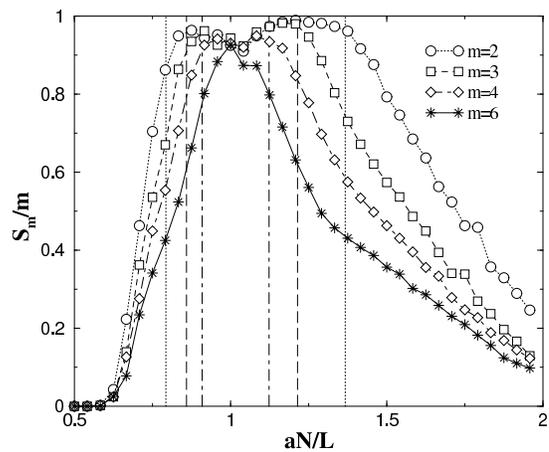}
    \caption{Entropy $S_m/m$ of the El Farol bar problem with true
    information and $L=60$, $\bar a=1/2$ and $m=2,3,4$ and $6$ as a
    function of $N$. Vertical lines delimit the symmetric phase of
    the corresponding model with random information.}
    \label{entropy}
\end{figure}

\section{True information}

The behavior of the model with true information deviates from that
with random information because, under the dynamics Eq. \req{trueinfo}, the
space of informations is not sampled uniformly. More precisely, if
$\rho^\mu$ is the probability of information $\mu$ in the stationary
state, we can make this statement quantitative introducing the entropy

\[
S_m=-\sum_{\mu=1}^{2^m}\rho^\mu\log_2\rho^\mu.
\]
When the space of information is sampled uniformly $\rho^\mu=1/2^m$ we
find $S_m=m$, whereas $S_m=0$ when only one value of $\mu$ is sampled
recursively. Fig. \ref{entropy} shows that $S_m\approx m$ only occurs
in the region where $H\simeq 0$ in the corresponding model with random
information, i.e. in the symmetric phase. This is consistent, because
if $H=0$ then the process of Eq. \req{trueinfo} is a simple diffusion
on the so-called De Bruijn graph. We refer the interested reader to
Ref. \cite{CM99} for a detailed account of this process. Here it
is sufficient to observe that if $\avg{A|\mu}=L$ for all $\mu$,
then\footnote{Notice that $A(t)$ is the sum of $N$ terms $\pm 1$,
which is asymptotically symmetric around the mean.}
\[
{\rm Prob}\{\mu(t+1)=|2\mu(t)+1|_{2^m}\}=\frac{1}{2}.
\]
If this is the case, the stationary state probability $\rho^\mu=1/2^m$
is uniform \cite{CM99}. When $H>0$, for a particular value of
$\mu$, we expect that $L-A(t)$ will take more frequently one sign or
the other. Hence Eq. \req{trueinfo} will induce a biased diffusion
process on $\mu(t)$. In particular, for $\ovl{a}N < L$, the attendance will
be more often below the comfort level than otherwise. This means that
$1$'s will occur more often that $0$'s in Eq. \req{mut} for $\mu(t)$.
It is easy to check that a systematic bias of this type, produces a
distribution $\rho^\mu$ which is concentrated on $\mu=\{111\ldots\}$
(using the binary representation). Likewise $\rho^\mu$ is peaked on
$\mu=\{000\ldots\}$ when $a N>L$~\footnote{As a side remark, if ${\rm
Prob}\{\mu(t+1)=|2\mu(t)+1|_{2^m}\}=p$, $\rho^\mu=p^{n(\mu)}(1-p)^{m-n(\mu)}$ where $n(\mu)$ is the number of 1 in the binary representation of $\mu$, and $-S(p)/m=p\log p +(1-p)\log(1-p)$}.

Hence outside the symmetric phase, when $H>0$, the process $\mu(t)$
acquires a bias, which reduces the ``effective number of information
patterns'' to a number $2^{S_m}$. In order to understand how this
changes the collective behavior of players, imagine the extreme case
$S_m=0$ where for some reason, the state $\mu(t)=\{000\ldots\}\equiv
0$ occurs for a large number of periods. Then agents will learn how to
respond optimally to this state $\mu=0$.

There are $N_{++}={\bar a}^2 N$ agents with
$a_{i,1}^0=a_{i,2}^0=1$. They will go anyway. There are
$N_{--}=(1-{\bar a})^2 N$ agents with $a_{i,1}^0=a_{i,2}^0=0$ who will
not go. The remaining $2\bar a(1-{\bar a})N$ can decide. 
Agents can learn to converge to $A(t)=L$ provided that $N_{++}\le L\le
N-N_{--}$, i.e. if
\[
\frac{L}{1-(1-{\bar a})^2}\le N\le \frac{L}{{\bar a}^2}.
\]
Furthermore, if $N>L/\bar a$, in particular if it is close to the
upper limit $L/\bar a^2$, the information $\mu=0$ will arise very
frequently from the dynamics Eq. \req{trueinfo}. The same argument
runs for $N<L/\bar a$ and it shows agents can coordinate quite
efficiently when $N$ is close to $L/[1(1-{\bar a})^2]$, because then
the information $\mu=\{111\ldots\}$ will almost always occur.

This complex interaction between the dynamics of $A(t)$ and $\mu(t)$
explains the non-monotonic behavior of $\avg{A}-L$ in Fig. \ref{fig1}. 

\section{Conclusions}

We have presented a complete theory of the El Farol bar problem.  The
key issue lies in the definition of the strategy space. First we have
shown that, for the most general ensemble of strategies, convergence
of the attendance to the comfort level is a trivial consequence of the
law of large numbers. It does not even require inductive rationality.
Even zero-intelligent agents are able to reach it.  This is likely to
be true for any reasonable predictor based strategy, in particular for
unbiased ones.

We further focus attention on a particular ensemble of strategies,
with a desirable continuity property. This leads us to study models
very similar to the Minority game. We first introduce the random
information version of the game, for which statistical physics
provides a complete theoretical understanding and analyze the
consequences of the dynamics of true information. 

It turns out that as the parameter of the El Farol bar problem change,
the system performs a trajectory on a phase diagram characterized by a
symmetric phase where $\avg{A}=L$ and a phase where $\avg{A}\not = L$.
Deep in the symmetric phase, anomalous fluctuations similar to crowd
effects, develop making the coordination of agents even worse than
that of random agents in some cases. It is precisely close to the
phase boundary that agents manage to coordinate most
efficiently. This, however, requires a small bias of either sign, in
the strategies of agents.

These findings not only confirm that the El Farol bar problem is
indeed a quite interesting complex system. But they also show that a
coherent understanding of its behavior is possible, using tools of
statistical physics.

\end{document}